\documentclass{elsart}%
\usepackage{amsmath}
\usepackage{amsfonts}
\usepackage{amssymb}
\usepackage{graphicx}%
\usepackage{fixmath}%
\begin{document}
\begin{frontmatter}
\title{$\kappa$-deformed Dirac oscillator in an external magnetic field}
\author[ya,yb]{Y. Chargui}
\author[ya,yb]{A. Dhahbi}
\author[ba]{B. Cherif}
\corauth[cor]{Corresponding author Email: yassine.chargui@gmail.com}
\address[ya]{Physics Department, College of Science and Arts at ArRass, Qassim University, PO Box 53, ArRass 51921,
Saudi Arabia}
\address[yb]{Universit\'{e} de Tunis El Manar, Facult{\'e} des Sciences de Tunis, Unit{\'e} de Recherche de Physique Nucl{\'e}aire et des Hautes Energies, 2092 Tunis, Tunisie}
\address[ba]{Department of Mathematics, College of Science and Arts at ArRass, Qassim University, PO Box 53, ArRass 51921,
Saudi Arabia}
\begin{abstract}
We study the solutions of the $(2+1)$-dimensional $\kappa$-deformed Dirac oscillator in the presence of a constant transverse magnetic field. We demonstrate how the deformation parameter affects the energy eigenvalues of the system and the corresponding eigenfunctions. Our findings suggest that this system could be used to detect experimentally the effect of the deformation. We also show that the hidden supersymmetry of the non-deformed system reduces to a hidden pseudo-supersymmetry having the same algebraic structure as a result of the $\kappa$-deformation.
\end{abstract}
\end{frontmatter}

\section{Introduction}
The Dirac oscillator system was considered for the first time by Ito \emph{et al} \cite{1}, and then reintroduced by Moshinsky and Szczepaniak \cite{2}, who called it by this name because its non-relativistic limit gives a harmonic oscillator with a very strong spin-orbit coupling term. At the beginning, the interest in this system was mainly motivated by its relevance to describe the confinement of quarks in QCD \cite{3,3b,3c}. But later, the Dirac oscillator has become a matter of study in different branches of physics. As a matter of fact, it has become a central component of mathematical physics where it provides a paradigm for covariant quantum models with well determined non-relativistic limit \cite{5,7,8,9,10,11,12,13,14,15,16,17,18}. Moreover, the Dirac oscillator appears in nuclear and sub-nuclear physics \cite{19,20,21,21b,21c}, quantum optics \cite{22,23,23c,24,25}, supersymmetry \cite{22b,23b,24b,25b,26,27}, non-commutativity \cite{28,29,30,31} and even in connection with Higgs symmetry \cite{32}. Additionally, it has been argued recently that the Dirac oscillator model can describe some properties of electrons in graphene \cite{graph}. Furthermore, the one-dimensional Dirac oscillator had recently found its first experimental realization, based on its relation to a corresponding tight-binding system \cite{expreal}. This has opened a new field of investigation concerning the experimental applications of this model. Besides, over the years, the Dirac oscillator has been well studied in the context of deformed algebras \cite{10,33,34,43,43b}. A particularly interesting quantum deformation theory is the one based on the $\kappa$-Poincar\'{e}-Hopf algebra \cite{35,36,37,38,39,40}. The latter is defined by the following commutation relations
\begin{eqnarray}\label{kdefalg}
\left[p_{\mu},p_{\nu}\right]=0\nonumber\\
\left[M_{i},p_{\mu}\right]=\left(1-\delta_{0\mu}\right)i\varepsilon_{ijk}p_{k}\nonumber\\
\left[L_{i},p_{\mu}\right]=i\left[p_{i}\right]^{\delta_{0\mu}}\left[\delta_{ij}\epsilon^{-1}
\sinh\left(\epsilon p_{0}\right)\right]^{1-\delta_{0\mu}}\nonumber\\
\left[M_{i},M_{j}\right]=i\varepsilon_{ijk}M_{k},\quad \left[M_{i},L_{j}\right]=i\varepsilon_{ijk}L_{k}\nonumber\\
\left[L_{i},L_{j}\right]=-i\varepsilon_{ijk}\left[M_{k}\cosh\left(\epsilon p_{0}\right)-\frac{\epsilon^{2}}{4}p_{k}p_{l}M_{l}\right]
\end{eqnarray}
where $p_{\mu}=\left(p_{0},\textbf{p}\right)$ are the deformed generators for energy and momenta, $M_{i}$ and $L_{i}$ are respectively spatial rotations and deformed boost generators. The group parameter $\epsilon$ is defined by
\begin{equation}\label{eps}
\epsilon=\kappa^{-1}=\lim_{R\rightarrow\infty}R\ln q
\end{equation}
with $R$ the de Sitter radius and $q$ a real deformation parameter. The deformed algebra $(\ref{kdefalg})$ implies a modified Dirac equation, usually called the $\kappa$-Dirac equation. Recently, this equation has been the focus of several works investigating the possible effects of the deformation on the physical properties of various relativistic quantum systems. In this context we can mention the study of the Dirac-Coulomb problem \cite{41}, the spin-$1/2$ Aharonov-Bohm effect \cite{41b,42}, the Dirac oscillator \cite{43,43b}, Landau levels and the integer quantum Hall effect \cite{44,44b} and the Dirac equation in crossed magnetic and electric fields \cite{45}. In this letter, we aim to study the effects of quantum deformations on the $(2+1)$-dimensional Dirac oscillator system, when a constant transverse magnetic field is applied. In the non-deformed case this problem admits an exact solution \cite{45b,45c}, and the model is known to be relevant for the description of many realistic physical situations. For instance, it was proposed as a relativistic extension of the parabolic non-relativistic harmonic potential used to describe quantum confined few-electrons systems in two dimensions \cite{46}. In addition, the exact solvability of this system was exploited to describe analytically several aspects of chiral quantum phase transition \cite{13}. We organize our work as follows. In the second section we study the first order effect of the deformation on solutions of the two-dimensional Dirac oscillator interacting with a transverse magnetic field. In the third section we address the question of the hidden supersymmetry of the system in connection with the introduced deformation. In the final section we give our conclusion.

\section{The $2D$ $\kappa$-Dirac oscillator in a magnetic field}
In this section, we undertake to solve the two-dimensional $\kappa$-Dirac oscillator, invariant under the $\kappa$-Poincar\'{e} quantum algebra, in the presence of a constant transverse magnetic field. We consider the situation when the electron is confined to move in the $xy$-plane and the external magnetic field $\textbf{B}$ acts along the $z$-direction. The underlying magnetic vector potential $\textbf{A}$ can be chosen in the symmetric gauge as $\textbf{A}=\left(-\frac{B}{2}y,\frac{B}{2}x,0\right)$. The problem is then independent of the $z$ variable. Therefore, our starting point will be the free $\kappa$-Dirac equation in $(2+1)$-dimensions which, up to the first order in $\epsilon$, reads \cite{37,38}:
\begin{equation}\label{kde}
\left\{\gamma_{0}p_{0}-\gamma_{i}p_{i}+\frac{\epsilon}{2}
\left[\gamma_{0}\left(p_{0}^{2}-p_{i}p_{i}\right)-mp_{0}\right]\right\}\psi=m\psi
\end{equation}
where $i=1,2$ and $\psi$ a two-component spinor. Moreover, we shall consider the following representation of the gamma matrices
\begin{equation}\label{gama}
\gamma_{0}=\beta=\sigma_{z},\quad\gamma_{1}=\beta\sigma_{x},\quad\gamma_{2}=s\beta\sigma_{y}
\end{equation}
where $\sigma_{i}$ are Pauli matrices and parameter $s$ is introduced to distinguish the two spin states, with $s=+1$ for spin"up" and $s=-1$ for spin "down". Now, the system we are considering stems from the prescriptions :
\begin{equation}\label{presc}
p_{0}\rightarrow H,\quad \mathbf{p}\rightarrow\mathbold{\pi}=\textbf{p}-e\mathbf{A}-im\omega\beta\mathbf{r}
\end{equation}
where $e$ is the Dirac oscillator, $m$ is its mass and $\omega$ stands for its angular frequency. For a stationary state with an energy $E$, these substitutions result in the following eigenvalue equation
\begin{equation}\label{hamileq}
H\psi=\left(\mathbold{\alpha}.\mathbold{\pi}\right)+m\beta\left(1+\frac{\epsilon}{2}H\right)
-\frac{\epsilon}{2}\left(H^{2}-\mathbold{\pi}^{2}\right)=E\psi
\end{equation}
with
\begin{equation}
\mathbold{\alpha}.\mathbold{\pi}=\sigma_{x}\pi_{x}+s\sigma_{y}\pi_{y}
\end{equation}
Then, by iterating the previous equation and retaining only terms of first order in $\epsilon$, the involved Hamiltonian reduces to the form
\begin{equation}\label{hamil}
H=\left(1+\frac{m\epsilon}{2}\beta\right)\left(\mathbold{\alpha}.\mathbold{\pi}\right)+m\beta
-\frac{\epsilon}{2}\left[\left(\mathbold{\alpha}.\mathbold{\pi}\right)^{2}-\mathbold{\pi}^{2}\right]
\end{equation}
It is worth noting here that, because of the term $\mathbold{\pi}^{2}$ induced by the deformation, the Hamiltonian of the system is no longer Hermitian, but it is $PT$-symmetric, i.e., invariant under $PT$ transformation where $P$ and $T$ are, respectively, the operators of parity and time-reversal transformations. Furthermore, we will show later that the radial Hamiltonian resulting from $(\ref{hamil})$ is pseudo-Hermitian. Besides, we can easily verify that
\begin{equation}
\left(\mathbold{\alpha}.\mathbold{\pi}\right)^{2}-\mathbold{\pi}^{2}=2m^{2}\omega\Omega r^{2}-2m\omega sL_{z}+2ms\omega_{c}\sigma_{z}+2im\omega\sigma_{z}\left(\mathbold{r}.\mathbold{p}\right)
\end{equation}
with the notation $\Omega=\omega-s\omega_{c}$ and $\omega_{c}=eB/m$ is the cyclotron frequency. Now we can readily check that the total angular momentum of the
system, $J_{z}=L_{z}+\frac{1}{2}\sigma_{z}$ commutes with the Hamiltonian $(\ref{hamil})$. So $J_{z}$ is conserved. Here $L_{z}$ is the angular momentum which can be represented as $L_{z}=-i\partial_{\theta}$.  Next, let us make the following ansatz for the total wave function
\begin{equation}
\psi\left(\overrightarrow{r}\right)=e^{-is\theta\sigma_{z}}\phi\left(r,\theta\right)
\end{equation}
where $\left(r,\theta\right)$ are polar coordinates. Then we can establish that
\begin{equation}
\left(\overrightarrow{\alpha}.\overrightarrow{\pi}\right)\psi=
-i\sigma_{x}\left(\partial_{r}\phi+\frac{\phi}{r}\right)-\frac{is}{r}\sigma_{y}\partial_{\theta}\phi-mr\Omega\sigma_{y}\phi
\end{equation}
and
\begin{equation}
L_{z}\psi=-e^{-is\theta\sigma_{z}}\left(s\sigma_{z}+i\partial_{\theta}\right)\phi
\end{equation}
This permits to rewrite the eigenvalue equation $(\ref{hamileq})$ under the form
\begin{eqnarray}
\left(1+\frac{m\epsilon}{2}\sigma_{z}\right)e^{is\theta\sigma_{z}}
\left[\sigma_{x}\left(\partial_{r}\phi+\frac{\phi}{r}\right)+\sigma_{y}\left(\frac{s}{r}\partial_{\theta}\phi-imr\Omega\phi\right)\right]=
\nonumber\\
+i\epsilon\left[ m^{2}\omega\Omega r^{2}-m\omega sL_{z}+m\left(\omega+s\omega_{c}\right)\sigma_{z}+im\omega\sigma_{z}\left(\overrightarrow{r}.\overrightarrow{p}\right)\right]\phi
\nonumber\\
i\left(E-m\sigma_{z}\right)\phi
\end{eqnarray}
Now, since the system preserves its angular symmetry in the presence of the deformation, we can still use the usual ansatz for the function $\phi$
\begin{equation}
\phi\left(r,\theta\right)=\frac{e^{im_{l}\theta}}{\sqrt{r}}\left(\begin{array}{c}
                                                            f\left(r\right)e^{is\theta} \\
                                                            ig\left(r\right)
                                                          \end{array}
\right)
\end{equation}
where $m_{l}=0,\pm 1,\pm 2, \pm 3,...$ is the angular momentum quantum number. This yields the following set of coupled radial differential equations
\begin{eqnarray}\label{syst}
\left(1+\frac{m\epsilon}{2}\right)\left[g'+\left(\frac{j}{r}-m\Omega r\right)g\right]=
\nonumber\\
\left[E-m+\epsilon\left(m^{2}\omega\Omega r^{2}-m\left[\Omega+\omega(j-1)\right]\right)\right]f
+\epsilon m\omega rf'
\nonumber\\
\left(1-\frac{m\epsilon}{2}\right)\left[f'+\left(m\Omega r-\frac{j}{r}\right)f\right]=
\nonumber\\
-\left[E+m+\epsilon\left(m^{2}\omega\Omega r^{2}-m\left[\omega(1+j)-\Omega\right]\right)\right]g
+\epsilon m\omega rg'
\end{eqnarray}
where we have introduced the notation $j=sm_{l}+\frac{1}{2}$. Decoupling the previous system yields the following second order differential equation for the upper component $f$
\begin{eqnarray}\label{feq}
f''\left(r\right)+2m^{2}\omega\epsilon rf'\left(r\right)
\nonumber\\
+\left[\mu_{j}^{2}-\frac{j(j-1)}{r^{2}}-m^{2}\Omega\left(\Omega-2\epsilon\omega E\right)r^{2}\right]f\left(r\right)=0
\end{eqnarray}
where all terms of higher orders in $\epsilon$ have been kept out. The symbol $\mu_{j}^{2}$ is used for the quantity
\begin{equation}
\mu_{j}^{2}=E^{2}-m^{2}+2m^{2}\epsilon \left(s\omega_{c}+\frac{\omega}{2}\right)+(2j+1)m\left(\Omega-\epsilon\omega E\right)
\end{equation}
We assume for now that $\Omega\neq 0$. The specific case for which $\Omega=0$ will be considered later. Then, introducing the change of variable
\begin{equation}
\rho=m\left|\Omega\right|\left(1-\frac{\epsilon\omega E}{\Omega}\right)r^{2}
\end{equation}
and making the ansatz
\begin{equation}
f\left(\rho\right)=e^{-\left(1+\frac{m\epsilon\omega}{\left|\Omega\right|}\right)\rho/2}\rho^{\frac{1+2|m_{l}|}{4}}\xi\left(\rho\right)
\end{equation}
we get for the function $\xi$ the following differential equation
\begin{equation}
\rho\xi''\left(\rho\right)+\left(1+\left|m_{l}\right|-\rho\right)\xi'\left(\rho\right)-a\xi\left(\rho\right)=0
\end{equation}
where we have used the notation
\begin{equation}
a=\frac{1+\left|m_{l}\right|}{2}-\frac{\mu_{j}^{2}}{4m\left|\Omega\right|}\left(1+\frac{\epsilon\omega E}{\Omega}\right)+\frac{\epsilon m\omega}{4\Omega}
\end{equation}
The solution of the previous equation that leads to a wave function regular at the origin is proportional to Kummer's function of the first kind $F\left(a,b,z\right)$
\begin{equation}
\xi\left(\rho\right)\propto F\left(a,1+\left|m_{l}\right|,\rho\right)
\end{equation}
This solution has an asymptotic behavior for large $|z|$ is dominated by
\begin{equation}\label{asybh}
F\left(a,b,z\right)\sim\Gamma(b)\left(\frac{z^{a-b}e^{z}}{\Gamma(a)}-\frac{(-z)^{-a}}{\Gamma(b-a)}\right)
\end{equation}
Therefore, in order to ensure the normalizability of the wave function $\psi$ we demand $a=-n$ with $n$ a positive integer. As a matter of fact,  $F\left(-n,b,z\right)$ is proportional to the generalized Laguerre polynomial of degree $n$ : $L_{n}^{b-1}\left(z\right)$. Thus, we can express the final solution obtained for $f$ as
\begin{equation}
f\left(\rho\right)=C_{n}e^{-\left(1+\frac{m\epsilon\omega}{\left|\Omega\right|}\right)\rho/2}\rho^{\frac{1+2|m_{l}|}{4}}
L_{n}^{\left|m_{l}\right|}\left(\rho\right)
\end{equation}
where $C_{n}$ is a normalization constant. In addition, the requirement $a=-n$ yields the following quantization condition for the energies :
\begin{equation}\label{ccdenf}
\left(E^{2}-m^{2}\right)\left(1+\frac{\epsilon\omega E}{\Omega}\right)=2m\left|\Omega\right| N_{s}-2\epsilon m^{2}s\omega_{c}
\end{equation}
with $ N_{s}=2n+|m_{l}|-sm_{l}$. Analogously, from system $(\ref{syst})$ we can also derive for the lower component $g$ an equation similar to Eq.$(\ref{feq})$ where the parameter $\mu^{2}_{j}$ and the number $j(j-1)$, coefficient of $1/r$, are replaced by $\mu^{2}_{j-1}$ and $j(j+1)$, respectively. Repeating the same steps followed in solving the equation of $f$, we find for $g$ the expression :
\begin{equation}
g\left(\rho\right)=C'_{n}e^{-\left(1+\frac{m\epsilon\omega}{\left|\Omega\right|}\right)\rho/2}\rho^{\frac{1+2|sm_{l}+1|}{4}}
L_{n}^{\left|1+sm_{l}\right|}\left(\rho\right)
\end{equation}
Moreover, we obtain the following energy equation
\begin{equation}\label{ccdeng}
\left(E^{2}-m^{2}\right)\left(1+\frac{\epsilon\omega E}{\Omega}\right)=2m\left|\Omega\right| N'_{s}-2\epsilon m^{2}s\omega_{c}
\end{equation}
with $ N'_{s}=2n+1+|sm_{l}+1|-sm_{l}$. Firstly we remark that the l.h.s in the two energy equations $(\ref{ccdenf})$ and $(\ref{ccdeng})$ is the same except for a constant $4m\Omega$. Later we will show that this property is related to the hidden pseudo-supersymmetry of the system. Notice also that the limit $\epsilon\rightarrow 0$ allows to recover the right results for the spectrum of the non-deformed two-dimensional Dirac oscillator with a transverse magnetic field \cite{46}. This reveals the consistency of our calculations. Furthermore, it is readily apparent from Eqs.$(\ref{ccdenf})$ and $(\ref{ccdeng})$ that the infinite degeneracy presented by the usual energies is preserved, exactly as in the case where the magnetic field is absent \cite{43b}. Thus we can deduce that the deformation does not break the underlying symmetry of the system. On the other hand, it is known that the presence of the transverse magnetic field does not change the dynamic of the non-deformed $(2+1)$-dimensional Dirac oscillator as long as $\Omega\neq 0$. It only changes its angular frequency from $\omega$ to $\Omega$ \cite{45b}. This observation remains valid in the presence of $\kappa$-deformation, but with a slight modification caused by the appearance of the additional term $-2\epsilon m^{2}s\omega_{c}$ in the r.h.s of Eqs.$(\ref{ccdenf})$ and $(\ref{ccdeng})$, which is a pure manifestation of the magnetic field.

Then, let us obtain the energy eigenvalues determined by Eq.$(\ref{ccdenf})$. By retaining only terms of first order in $\epsilon$, we find
\begin{equation}
E_{N_{s}}=\pm\sqrt{m^{2}+2m\vert\Omega\vert N_{s}-2\epsilon m^{2}s\omega_{c}}-\epsilon m\omega N_{s}
\end{equation}
We can see here that the presence of parameter $\epsilon$ makes the particle and antiparticle energies different. As explained in Ref.\cite{43}, this reveals the breaking of the charge conjugation symmetry of the system caused by the $\kappa$-deformation. However, this aspect of the energy spectrum has no relation with the coupling to the magnetic field. It is a characteristic response of the Dirac oscillator when the deformation is introduced.

Next, let us discuss the specific case for which $\Omega=0$. This case corresponds to spin up (down) states coupled to a magnetic field pointing up (down) having the special strength $B=m\omega/e$. In this situation, the differential equation obeyed by the upper component $f$ can be directly deduced from Eq.$(\ref{feq})$ and reads
\begin{eqnarray}\label{feqsc}
f''\left(r\right)+2m^{2}\omega\epsilon rf'\left(r\right)+\left[\mu_{0j}^{2}-\frac{j(j-1)}{r^{2}}\right]f\left(r\right)=0
\end{eqnarray}
where the notation $\mu_{0j}^{2}$ stands for $\mu_{j}^{2}\left(\Omega=0\right)$. We can easily show that the solution of the above equation, leading to a well-behaved wave function at the origin, is such that
\begin{equation}
f\left(r\right)\propto \rho^{\frac{1+2|m_{l}|}{4}}F\left(a_{0},1+\left|m_{l}\right|,\rho\right)
\end{equation}
with
\begin{equation}
\rho=-m^{2}\omega\epsilon r^{2},\quad a_{0}=\frac{\mu_{0j}^{2}}{4m^{2}\omega\epsilon}+\frac{1}{4}+\frac{\left|m_{l}\right|}{2}
\end{equation}
Furthermore, given the asymptotic behavior $(\ref{asybh})$ of Kemmer's function for large $|z|$, we should impose the condition $b-a=-n$ with $n$ a positive integer, in order to ensure an exponential decay of the bound-state wave function for large $r$. This requirement entails the following energy spectrum :
\begin{equation}\label{espcrit}
E_{n}=\pm m\left[1+\omega\epsilon\left(2n+|m_{l}|\pm m_{l}\right)\right]
\end{equation}
Thus we can observe that, in this specific situation, the $\kappa$-deformation changes drastically the dynamics of Dirac oscillator in comparison with its dynamics in the absence of the deformation. In the latter case the Dirac oscillator stop oscillating when the specific magnetic field is applied \cite{45b}. This behaviour might be deduced from $(\ref{espcrit})$ by putting $\epsilon=0$. However, expression $(\ref{espcrit})$ shows that the presence of the deformation entails slight oscillations of the system which acquires a quantized energy just as a non-relativistic harmonic oscillator. In our opinion, this result suggests that the effect of deformation could be detected in an experiment based on the situation we have been considering, especially given recent advances made in the experimental realization of the Dirac oscillator system.
\section{Hidden pseudo-supersymmetry}
Firstly, we recall that, in the non-deformed case, it has been shown that the radial equations of the Dirac oscillator exhibits a supersymmetric character \cite{22b,25b}. In this section, we shall investigate the effect of the $\kappa$-deformation of the system on this hidden supersymmetry. For this purpose we will bring Eqs.$(\ref{syst})$ to a more relevant form. Multiplying system $(\ref{syst})$ from the left by the matrix $\left(1-\sigma_{z}\frac{m\epsilon}{2}\right)$ and defining the radial vector-function $\varphi\left(r\right)=\left(\begin{array}{cc}
                                                                                                   f\left(r\right) &, ig\left(r\right)
                                                                                                 \end{array}
\right)^{t}$, we can easily establish the following eigenvalue equation
\begin{equation}\label{radeq}
E_{\epsilon}\varphi\left(r\right)=H_{r}\varphi\left(r\right)
\end{equation}
where $E_{\epsilon}=E+\frac{m^{2}\epsilon}{2}$ and the radial Hamiltonian $H_{r}$ is given by
\begin{eqnarray}
H_{r}=\left(\sigma_{x}-i\epsilon m\omega r\sigma_{z}\right)p_{r}+M_{\epsilon}\sigma_{z}-K(r)\sigma_{y}
+\epsilon m\omega\left(j-m\Omega r^{2}\right)
\end{eqnarray}
with the notations
\begin{eqnarray}
p_{r}=-i\partial_{r},\quad M_{\epsilon}=m\left(1+\frac{\epsilon E}{2}-\epsilon s\omega_{c}\right),\quad K(r)=m\Omega r-j/r
\end{eqnarray}
Now the operation on Eq.$(\ref{radeq})$ from the left by $1-\epsilon m\omega r\sigma_{y}$ has the only effect of transforming the radial Hamiltonian to the form
\begin{eqnarray}
\widetilde{H}_{r}=\left(p_{r}-i\epsilon m^{2}\omega r\right)\sigma_{x}+\left(\epsilon m\omega E r-K(r)\right)\sigma_{y}+M_{\epsilon}\sigma_{z}
\end{eqnarray}
It should be noted here that $\widetilde{H}_{r}$ is pseudo-Hermitian \cite{pseudo}. Indeed, it is not hard to check that this Hamiltonian satisfy the following property
\begin{equation}
\eta_{\varepsilon}\widetilde{H}_{r}\eta_{\epsilon}^{-1}=\widetilde{H}_{r}^{\dag}
\end{equation}
where $\eta$ is the linear Hermitian automorphism defined as $\eta_{\epsilon}=\textbf{1}e^{2im^{2}\epsilon\omega r\partial _{p_{r}}}$ with $\textbf{1}$ the $2\times2$ identity matrix, and the subscript $"\dag"$ refers to the Hermitian adjoint of the operator. We can see here that this property of pseudo-Hermiticity of the Hamiltonian is a consequence of the $\kappa$-deformation of system. In the standard case ($\epsilon=0$), we have $\eta_{\epsilon}=\textbf{1}$ and $\widetilde{H}_{r}$ is Hermitian. Besides, the operator $\widetilde{H}_{r}$ can be written in a matrix form as
\begin{equation}
\widetilde{H}_{r}=\left(
                          \begin{array}{lr}
                            M_{\epsilon}  & A_{1} \\
                            A_{2}  & -M_{\epsilon} \\
                          \end{array}
                        \right)
\end{equation}
where $A_{1}$ and $A_{2}$ are the operators defined as
\begin{eqnarray}
A_{1}=p_{r}+i\left[K(r)-\epsilon\left(E+m\right)m\omega r\right],
\nonumber\\
A_{2}=p_{r}-i\left[K(r)-\epsilon\left(E-m\right)m\omega r\right]
\end{eqnarray}
Furthermore, we can readily verify that $A_{1}$ is the adjoint pseudo-hermitian of $A_{2}$ and vice versa. This means that
\begin{eqnarray}\label{psher}
A_{1}^{\sharp}=A_{2},\quad A_{2}^{\sharp}=A_{1},\quad \text{with} \quad O^{\sharp}=\eta_{\varepsilon}^{-1}O^{\dag}\eta_{\epsilon}
\end{eqnarray}
Now, the two components of $\varphi$ can be decoupled by calculating $\widetilde{H_{r}}^{2}$ and using the property $(\ref{psher})$ to rewrite the eigenvalue equation as
\begin{eqnarray}\label{eeqsusy}
\left(E_{\epsilon}^{2}- M_{\epsilon}^{2}\right)\varphi\left(r\right)=\left(
                          \begin{array}{lr}
                              A_{1}A_{1}^{\sharp}&0 \\
                            0  & A_{1}^{\sharp}A_{1} \\
                          \end{array}
                        \right)\varphi\left(r\right)=\left(
                          \begin{array}{lr}
                              H_{+}&0 \\
                            0  &  H_{-} \\
                          \end{array}
                        \right)\varphi\left(r\right)
\end{eqnarray}
In the above equation $ H_{+}$ and $ H_{-}$ are pseudo-supersymmetric partners \cite{pseudo}. The generators of the pseudo-supersymmetry transformations are the pseudo-supercharges $Q$ and $Q^{\sharp}$ defined as
\begin{equation}
Q=\left(
    \begin{array}{cc}
      0 &  A_{1} \\
      0 & 0 \\
    \end{array}
  \right),\quad Q^{\sharp}=\left(
    \begin{array}{cc}
      0 &  0\\
      A_{1}^{\sharp} & 0 \\
    \end{array}
  \right),\quad
\end{equation}
These pseudo-supercharges, together with the operator $H^{2}_{s}:=\widetilde{H_{r}}^{2}- M_{\epsilon}^{2}$, satisfy an $S(2)$ superalgebra
\begin{equation}
\left\{Q,Q^{\sharp}\right\}=H^{2}_{s},\quad \left[Q,H^{2}_{s}\right]=\left[Q^{\sharp},H^{2}_{s}\right]=0
\end{equation}
Thus, we see that due to the $\kappa$-deformation, the hidden supersymmetry of two-dimensional Dirac oscillator, previously discussed in the literature \cite{25b}, reduces to a hidden pseudo-supersymmetry having the same algebraic structure. This pseudo-supersymmetry is responsible of the similarity between energy equations obtained for the upper and the lower components of the radial function $\varphi$. Notice also that, as in the non-deformed case, this symmetry is not altered by the presence of the transverse magnetic field.

\section{Conclusion}
In conclusion, this work has investigated the first order effect of the $\kappa$-deformed Poincar\'{e}-Hopf algebra on the dynamics of the $(2+1)$-dimensional Dirac oscillator coupled to a transverse constant magnetic field. Our findings demonstrate how the energy eigenvalues of the system and the associated eigenfunctions are affected by the deformation parameter. In particular we have found that the deformation preserves the infinite degeneracy presented by the energies of the non-deformed system. Nevertheless its presence engenders an asymmetry between particle and antiparticle energy spectra. This fact reveals the $\kappa$-breaking of the charge conjugation symmetry of the usual system. We have also discussed the specific case where the angular frequency of the Dirac oscillator is equal to the cyclotron frequency. Here we have observed that the dynamics of the deformed system is completely different form its dynamics without deformation. In this situation, while the usual Dirac oscillator ceases moving, the $\kappa$-Dirac oscillator carries on slightly oscillating with a quantized energy just as a non-relativistic harmonic oscillator. Besides, we have examined the effect of the deformation on the supersymmetric character known for the radial equations of the Dirac oscillator. In this respect we concluded that this supersymmetry reduces to a pseudo-symmetry as a result of the $\kappa$-deformation.

\section*{Acknowledgments}
The authors gratefully acknowledge Qassim University, represented by the Deanship of Scientific Research, on the material support for this research under the number (1968-alrasscac-2016-1-12-S) during the academic year 1437 AH/2016 AD

\end{document}